\documentclass[aps,prl,twocolumn,superscriptaddress]{revtex4-2}

\usepackage{graphicx}
\usepackage{color}

\newcommand{\overbar}[1]{\mkern 1.5mu\overline{\mkern-1.5mu#1\mkern-1.5mu}\mkern 1.5mu}

\usepackage{mhchem}

\graphicspath {{figures/}}

\begin{document}

\title{Current driven insulator-to-metal transition without Mott breakdown in Ca$_2$RuO$_4$}

\author{Davide Curcio}
\affiliation{Department of Physics and Astronomy, Aarhus University, 8000 Aarhus C, Denmark}

\author{Charlotte E. Sanders}
\affiliation{Central Laser Facility, STFC Rutherford Appleton Laboratory, Harwell OX11 0QX, United Kingdom}

\author{Alla Chikina}
\affiliation{Department of Physics and Astronomy, Aarhus University, 8000 Aarhus C, Denmark}

\author{Henriette E. Lund}
\affiliation{Department of Physics and Astronomy, Aarhus University, 8000 Aarhus C, Denmark}

\author{Marco Bianchi}
\affiliation{Department of Physics and Astronomy, Aarhus University, 8000 Aarhus C, Denmark}

\author{Veronica Granata}
\affiliation{Dipartimento di Fisica 'E.R. Caianiello', Universit\'a degli Studi di Salerno, via Giovanni Paolo II 132, I-84084 Fisciano (Sa), Italy}

\author{Marco Cannavacciuolo}
\affiliation{Dipartimento di Fisica 'E.R. Caianiello', Universit\'a degli Studi di Salerno, via Giovanni Paolo II 132, I-84084 Fisciano (Sa), Italy}

\author{Giuseppe Cuono}
\affiliation{International Research Centre Magtop, Institute of Physics, Polish Academy of Sciences, Aleja Lotnik\'ow 32/46, PL-02668 Warsaw, Poland}

\author{Carmine Autieri}
\affiliation{International Research Centre Magtop, Institute of Physics, Polish Academy of Sciences, Aleja Lotnik\'ow 32/46, PL-02668 Warsaw, Poland}

\author{Filomena Forte}
\affiliation{CNR-SPIN, via Giovanni Paolo II 132, I-84084 Fisciano, Italy}

\author{Alfonso Romano}
\affiliation{Dipartimento di Fisica 'E.R. Caianiello', Universit\'a degli Studi di Salerno, via Giovanni Paolo II 132, I-84084 Fisciano (Sa), Italy}

\author{Mario Cuoco}
\affiliation{CNR-SPIN, via Giovanni Paolo II 132, I-84084 Fisciano, Italy}

\author{Pavel Dudin}
\affiliation{Synchrotron SOLEIL, Gif-sur-Yvette, France}

\author{Jose Avila}
\affiliation{Synchrotron SOLEIL, Gif-sur-Yvette, France}

\author{Craig Polley}
\affiliation{MAX IV Laboratory, Lund University, Lund, Sweden}

\author{Thiagarajan Balasubramanian}
\affiliation{MAX IV Laboratory, Lund University, Lund, Sweden}

\author{Rosalba Fittipaldi}
\affiliation{CNR-SPIN, via Giovanni Paolo II 132, I-84084 Fisciano, Italy}

\author{Antonio Vecchione}
\affiliation{CNR-SPIN, via Giovanni Paolo II 132, I-84084 Fisciano, Italy}

\author{Philip Hofmann}
\email{philip@phys.au.dk}
\affiliation{Department of Physics and Astronomy, Aarhus University, 8000 Aarhus C, Denmark}

\begin{abstract}
The electrical control of a material's conductivity is at the heart of modern electronics. Conventionally, this control is achieved by tuning the density of mobile charge carriers. A completely different approach is possible in Mott insulators such as \ce{Ca2RuO4}, where an insulator-to-metal transition (IMT) can be induced by a weak electric field or current \cite{Cao:2020vr}. This phenomenon has numerous potential applications in, \textit{e.g.}, neuromorphic computing \cite{Liu:2012ah,Pickett:2013aa,Kumar:2017ab,Valle:2019aa,Valle:2021un}. While the driving force of the IMT is poorly understood, it has been thought to be a breakdown of the Mott state. Using \emph{in operando} angle-resolved photoemission spectroscopy, we show that this is not the case:  The current-driven conductive phase arises with only a minor reorganisation of the Mott state. This can be explained by the co-existence of structurally different domains  that emerge during the IMT \cite{Bertinshaw:2019aa,Cirillo:2019aa,Zhao:2019ux,Jenni:2020vk,Okazaki:2020wa}. Electronic structure calculations show that the boundaries between domains of slightly different structure lead to a drastic reduction of the overall gap.  This permits an increased conductivity, despite the persistent presence of the Mott state. This mechanism represents a paradigm shift  in the understanding of IMTs, because it does not rely on the simultaneous presence of a metallic and an insulating phase, but rather on the combined effect of structurally inhomogeneous Mott phases.
 \end{abstract}
 \maketitle

 An electrical current can be expected to trigger an insulator-to-metal transition (IMT) in many correlated oxides, since Joule heating can push the local temperature high enough for a temperature-induced transition to take place. However, in some materials, the IMT can be triggered by surprisingly low current densities, and the effect is thought to be mainly electronic \cite{Cao:2020vr}.
 
 A particularly well-established example is \ce{Ca2RuO4}, a Mott insulator with a complex phase diagram and a strong susceptibility to phase changes that can be triggered not only by electric fields and currents \cite{Nakamura:2013aa,Okazaki:2013aa,Sakaki:2013vm}, but also by temperature \cite{Alexander:1999aa,Jung:2003aa}, pressure \cite{Nakamura:2002ab}, doping \cite{Nakatsuji:2000ab,Miyashita:2021vr} and strain \cite{Ricco:2018aa}.
 The current-induced IMT in this system has been investigated using a large variety of tools.
 While the results are not generally consistent, owing to the complexity of the phase diagram and the still disputed role of Joule heating \cite{Zhang:2019ac,Fursich:2019aa,Mattoni:2020ab,Chiriaco:2020ve}, there are several established experimental facts: (1) The transition's fingerprint is a region of negative differential resistance in transport measurements \cite{Sakaki:2013vm,Okazaki:2013aa,Zhang:2019ac,Cirillo:2019aa}; (2) macroscopically, the transition proceeds across the sample starting from the negative electrode \cite{Zhang:2019ac,Chiriaco:2020ve}; and
 (3) the IMT is strongly coupled to a structural phase transition involving an expansion of the crystal's $c$ axis:
 Starting from the so-called ``S-phase,'' with a short $c$ axis, a fully metallic ``L-phase'' emerges, with a significant $c$ axis expansion.
 Diffraction techniques confirm the trend of $c$-axis expansion, accompanied by the simultaneous presence of different structural phases \cite{Bertinshaw:2019aa,Cirillo:2019aa,Zhao:2019ux,Jenni:2020vk,Okazaki:2020wa}. Joule heating might play a role in creating some of these \cite{Jenni:2020vk}.

 Manifestly absent from the experimental results are direct characterisations of the electronic structure in the current-induced phase.
 Such information can, in principle, be obtained from angle-resolved photoemission spectroscopy (ARPES), but the application of photoemission techniques is hindered by a simple experimental limitation:
 in a current-carrying sample, the voltage drop inside the sample area illuminated by the ultraviolet (UV) light spot for photoemission is sufficiently large to completely deteriorate the energy resolution of the experiment.
 Here we solve this problem by using a very small light spot, on the order of a few $\mu$m, allowing us to characterise the electronic structure throughout the IMT.
 In addition to spectroscopic information, this approach also gives access to the local potential landscape on the surface, and we can thus use it as a local transport measurement \cite{Curcio:2020aa,Hofmann:2021tj}.
 
 \begin{figure}
   \includegraphics[width=0.5\textwidth]{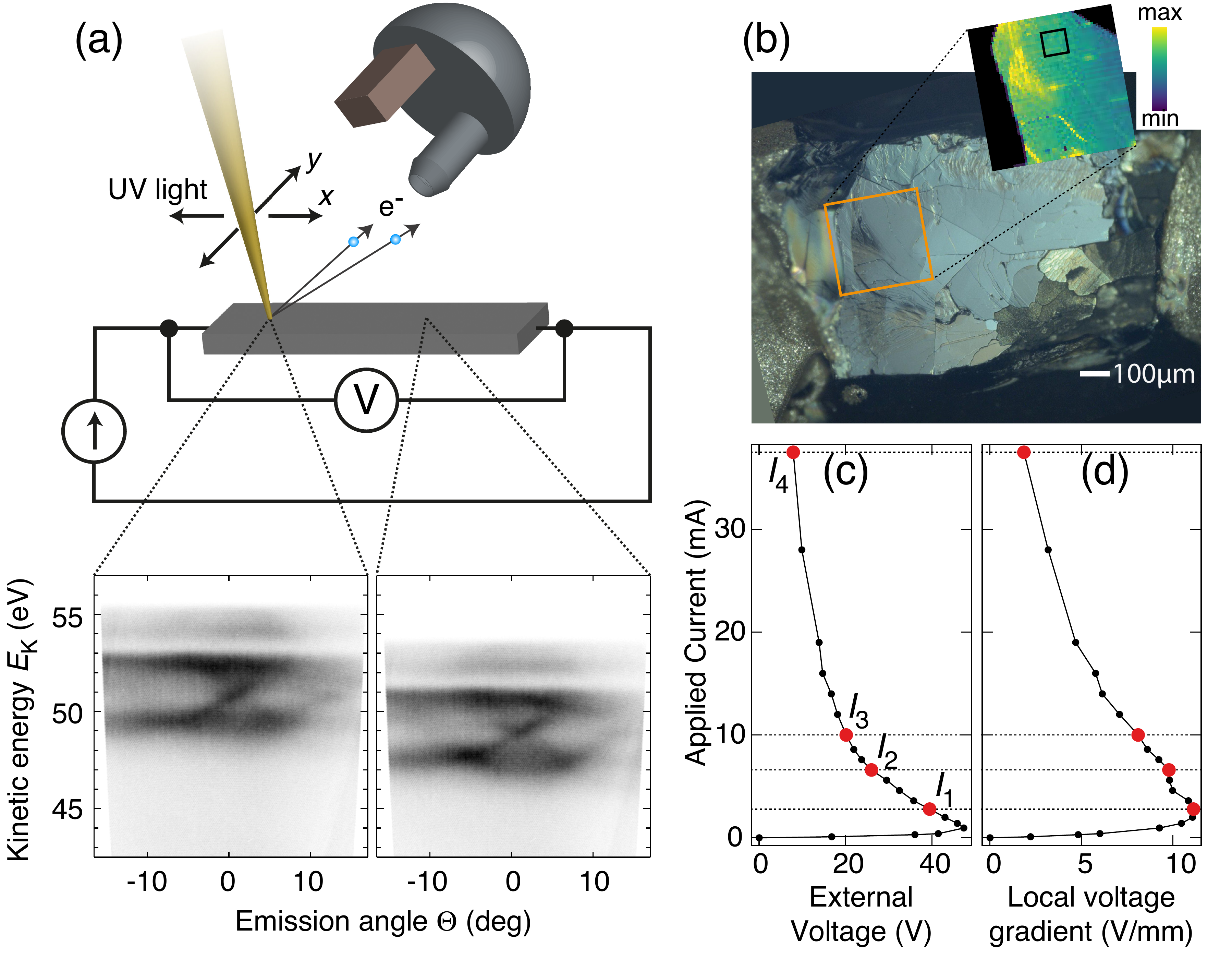}\\
   \caption{(a) Sketch of the experiment.
     The sample is exposed to a transport current in the $ab$ plane.
     The resulting voltage drop leads to a continuous energy shift of all states between the two electrodes.
     ARPES measurements are performed using a highly focused UV beam that is scanned across the surface, giving local spectra (photoemission intensity as a function of energy and emission angle) that are displaced in energy, following the local potential landscape across the sample.
     (b) Optical microscope image of the sample after the experiment.
     The contacts are seen on the left and right hand side.
     The area explored by ARPES is marked by an orange square.
     The inset shows a map of the photoemission intensity inside this area.
     (c) Externally measured $I/V$ curve and (d) local $I/V$ curve obtained from averaging the observed energy shift between the spectra within the black square marked in the inset of panel (b) (see Methods).
     The red markers indicate the current values for which data is displayed in Figures \ref{fig:2} and \ref{fig:3}.
   }
   \label{fig:1}
 \end{figure}
 
 The setup of the experiment is shown in Fig.~\ref{fig:1}(a).
 The sample is electrically contacted from the sides, so that both the electronic structure and the current-induced potential landscape can be mapped by the scanning of a highly focused UV beam across the sample surface.
 When this is done as a function of the applied current, it results in a five-dimensional data set:  the photoemission intensity as a function of the spatial coordinates $x$ and $y$, kinetic energy $E_\mathrm{k}$, emission angle $\Theta$ and the current $I$.
 Two local photoemission spectra (photoemission intensity as a function of $E_\mathrm{k}$ and $\Theta$) are given as examples.
 Due to the presence of the electric field accompanying the current density, the spectra collected at different points are displaced in energy.
 Fig.~\ref{fig:1}(b) shows an optical microscopy image of the sample used in this experiment. The current-dependent electronic structure has been studied in the area outlined by the orange square. The inset of Fig.~\ref{fig:1}(b) shows a map of the integrated photoemission intensity in this region, with intensity changes that clearly correspond to the sample's microscopic morphology. 
 
 Figure~\ref{fig:1}(c) shows the macroscopic $I/V$ curve, obtained from the applied current and the voltage drop across the entire sample.
 This is compared to a local $I/V$ curve (Fig.~\ref{fig:1}(d)) derived from the potential gradient measured by ARPES (see Methods) \cite{Curcio:2020aa}.
 Both show the shape typical for the current-induced IMT, with a pronounced region of negative differential resistance \cite{Sakaki:2013vm,Okazaki:2013aa,Zhang:2019ac,Cirillo:2019aa}.
 Note that the local $I/V$ curve is equivalent to a genuine four-point measurement and is thereby different from the external measurement, which includes the sample's contact resistance. A comparison of the curves requires the assumption that the ratio of total current and the current through the area probed by the local measurement is independent of the magnitude of the total current.
 The similarity of the two $I/V$ curves is essential for the interpretation of the spectroscopic data below:
 It shows that the local spectroscopic measurements reported here are representative of a sample area undergoing the IMT.

 \begin{figure}
   \includegraphics[width=0.48\textwidth]{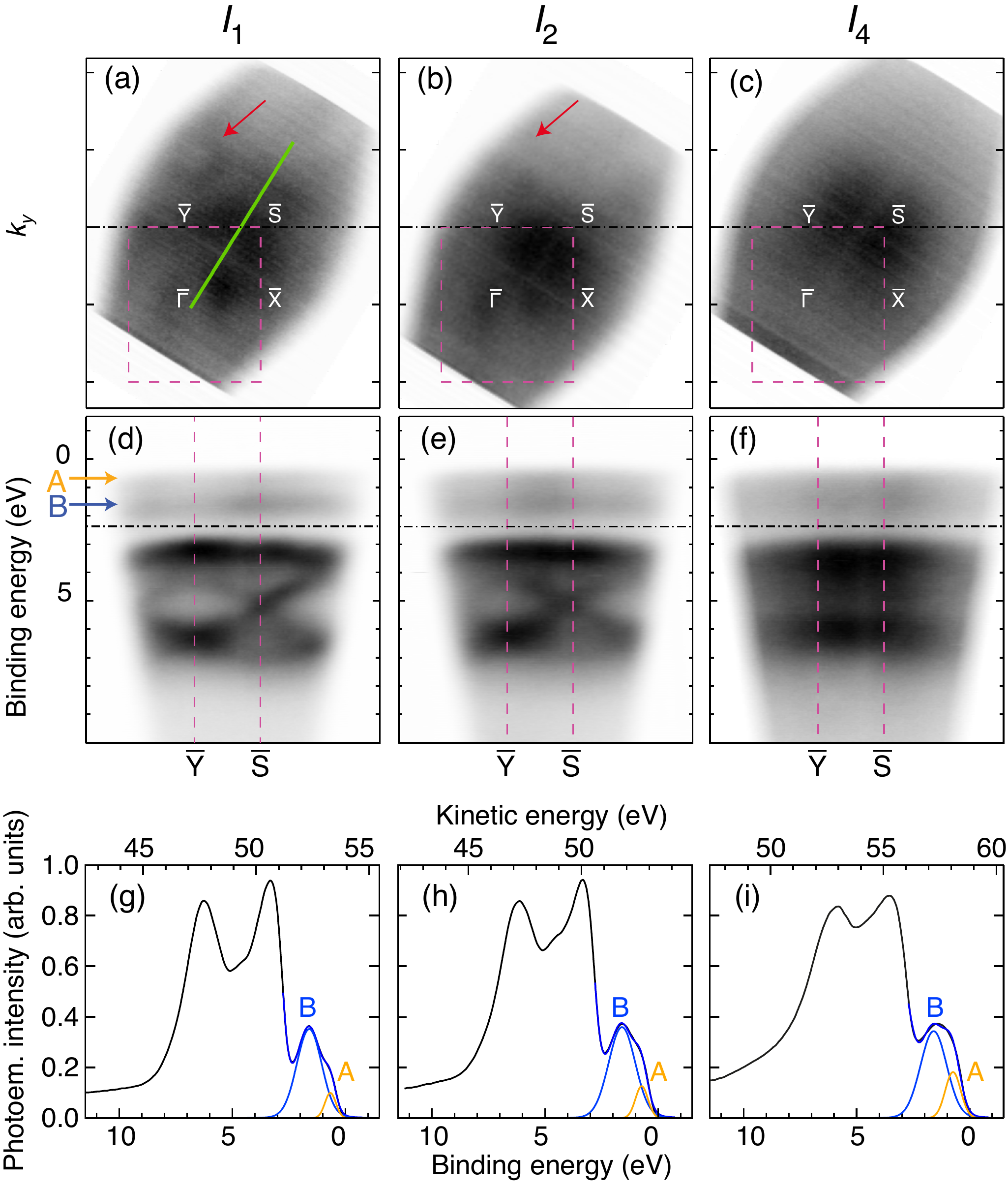}\\
   \caption{Spectral changes accompanying the IMT: (a)-(c) Photoemission intensity at a binding energy of 2.37~eV for $I_1$, $I_2$, $I_4$ (see markers on Fig.~\ref{fig:1}(c)).
     The surface Brillouin zone is superimposed.
     The green line marks the direction and range of the spectra in the multidimensional data set and in Fig.~\ref{fig:3}.
     The red arrow marks a circular constant energy contour stemming from the $d_{xy}$ states.
     (d)-(f) Intensity as a function of energy along the $\overbar{Y}\overbar{S}$ direction.
     (g)-(i) Angle-integrated energy distribution curves derived from data at $I_1$, $I_2$, $I_4$, respectively, integrated along the path identified by the green line in (a).
     The photoemission intensity at low binding energy has been fitted by two peaks, A and B, in the same way as in Ref.~\cite{Sutter:2017aa}.
     The fit components are shown and the positions are also marked in panels (d)-(f).}
   \label{fig:2}
 \end{figure}
 
 \begin{figure*}
   \includegraphics[width=0.9\textwidth]{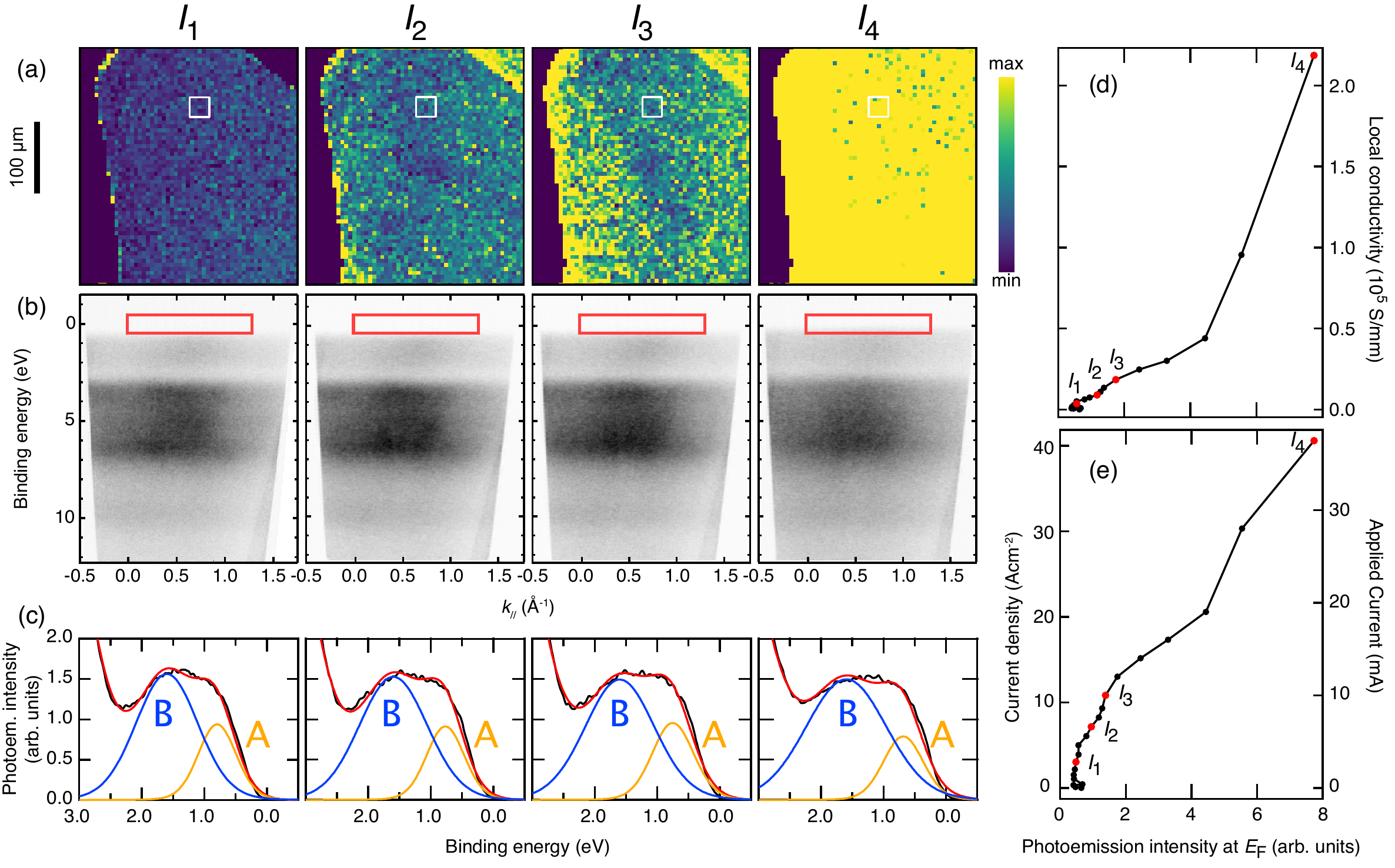}\\
   \caption{Mapping the IMT throughout the data set.
     (a) In-gap photoemission intensity across the sample for selected currents marked in Fig.~\ref{fig:1}(c).
     The integration region is given by the red rectangles in panel (b).
     (b) Corresponding photoemission spectra at the location marked by a white rectangles in panel (a).
     (c) Angle-integrated energy distribution curves obtained from the spectra in panel (b).
     (d) Estimated local conductivity and (e) applied current as a function of in-gap photoemission intensity, integrated over the red rectangles in panel (b).}
   \label{fig:3}
 \end{figure*}
 
The spectroscopic changes accompanying the IMT are illustrated in Fig.~\ref{fig:2} by measurements of the photoemission intensity throughout the Brillouin zone for three selected currents; (for additional data see Appendix).
 The low-current spectrum at $I_1$ is taken close to the turning point of the $I/V$ curve, \emph{i.e.}, close to the highest electric field, demonstrating the feasibility of the approach:
 Despite the field, the spectrum agrees very well with our own and  published equilibrium results \cite{Mizokawa:2001aa,Sutter:2017aa,Ricco:2018aa}.
 The two flat bands at low binding energies (marked by arrows) are related to the d$_{zx}$ and d$_{yz}$ states of the Ru-4d t$_{2g}$ manifold.
 In the insulating S-phase, these are half-filled, and are a manifestation of the lower Hubbard band in a Mott state driven by Coulomb interactions \cite{Sutter:2017aa}.
 The remaining d$_{xy}$ band is fully occupied, and thus insulating.
 It gives rise to the circular intensity feature marked by arrows in Fig.~\ref{fig:2}(a) and (b).
 The more intense bands at higher binding energy are primarily derived from oxygen states.
 
 The IMT to a conducting L-phase is predicted to drastically change the electronic structure, resulting in a metallic state with a large Fermi surface \cite{Petocchi:2021wy}.
 Clearly, this is at odds with the ARPES results, which show an absence of Fermi level crossings  at the higher currents $I_2$ and $I_4$ (see Fig.~\ref{fig:2}).
 Indeed, an increased current only results in minor changes of the spectra.
 The signature of the Mott state remains and the most prominent effect appears to be a mere $k$-broadening that washes out the dispersive features.
 This is particularly evident in the oxygen bands, but it also takes place in the Ru states.
 The broadening can either signal that $k_{\parallel}$ ceases to be a good quantum number, possibly because of the creation of defects or small crystalline domains, or it can point towards the simultaneous existence of structural domains with slightly different electronic structures, such that their incoherent superposition is observed here..
 Both mechanism are consistent with the observation of a broad distribution of structural modifications in diffraction experiments \cite{Bertinshaw:2019aa,Cirillo:2019aa,Zhao:2019ux,Jenni:2020vk,Okazaki:2020wa} (see Appendix for X-ray diffraction results from the crystals used here). 
 The persistence of the Mott state in all of these structural modifications is a key result of this paper.
 
 \begin{figure}
   \includegraphics[width=0.95\columnwidth]{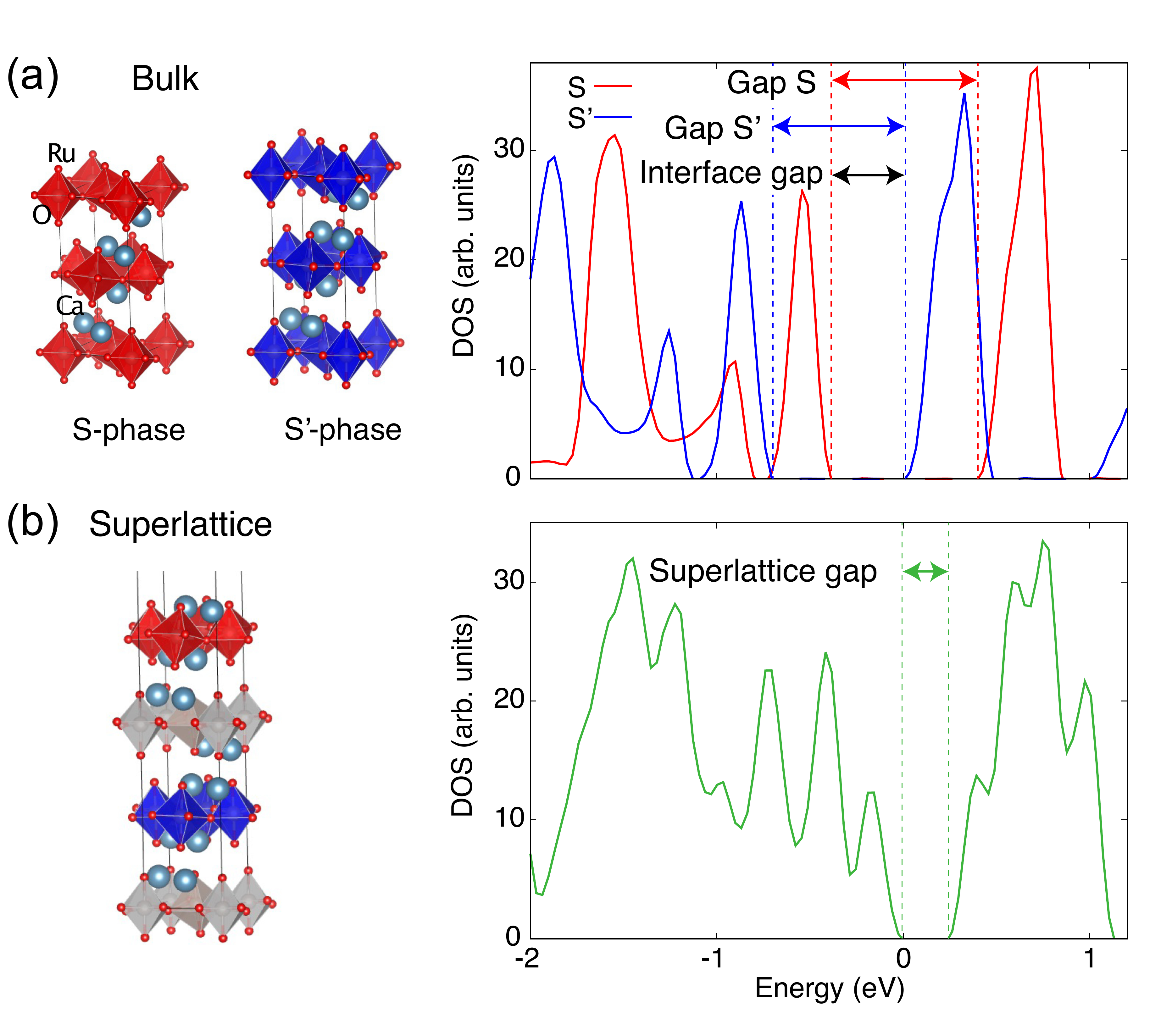}
   \caption{(a) Sketch of the \ce{Ca2RuO4} bulk crystal structures and corresponding density of states (DOS).
   S- and S$^{'}$ have unit cells with a  different $c$ axis length.
   S and S$^{'}$ are insulating with the indicated gaps but the band alignment leads to a smaller effective gap at the interface between S and S$^{'}$. 
   (b) Superlattice formed by the S and S$^{'}$ unit cells separated by an interface unit cell with intermediate properties (grey octahedra).
   The corresponding DOS shows a substantial reduction of the charge gap. The zero of the energy scale is fixed to the valence band maximum of the superlattice throughout the paper.
   }
   \label{fig:4}
 \end{figure}
 
 An important question is to what degree spectral intensity is filled into the Mott gap, triggering the transition in conductivity.
 In ARPES studies of superconductors or charge density systems, this can be answered very precisely by aligning all spectra to the Fermi energy $E_\mathrm{F}$ of the sample or to that of a metal in contact with the sample \cite{Lou:2022aa}.
 Here, this is not possible, due to the field-induced shifts between the spectra, which completely eliminate a common reference energy.
For a removal of these energy shifts, we first reduce the complexity of the spectra, by integrating each of them along the emission angle, resulting in angle-integrated energy distribution curve (EDC), as shown in Fig.~\ref{fig:2}(g)-(i).  The integration in angle is justified by the very small dispersion observed, especially in the \{d$_{zx}$,d$_{yz}$\} bands.
 Each EDC is fitted by two Gaussian peaks in the \{d$_{zx}$,d$_{yz}$\} energy region, called A and B \cite{Sutter:2017aa}.
 These fits are also shown in Figures \ref{fig:2}(g)-(i).
 We then choose a common energy reference point.
 The A peak is clearly unsuitable because it is closest to the gap where spectral changes are expected.
 Instead, we use the B peak, which appears to be a good choice for two reasons: (1) It represents a flat band in the equilibrium state; and (2) in contrast to the A peak, it is only weakly affected by the corresponding strain-induced IMT \cite{Ricco:2018aa}.
 Therefore, we align the B peak energy in all the spectra 78141 spectra of the now four-dimensional data set.
 Combining this with the energy difference between the B peak and $E_\mathrm{F}$ in equilibrium (1.60~eV, measured against a Fermi edge of polycrystalline Au), we obtain the local $E_\mathrm{F}$ for every spectrum.
 The spectra in Fig.~\ref{fig:2} are already aligned in this way and are shown as a function of binding energy.
 The original kinetic energy is also given for panels (g)-(i).
 
 The energy alignment of the spectra now opens the possibility to study the IMT throughout the entire dataset.
 To this end, Fig.~\ref{fig:3}(a) shows the hallmark of the IMT, the in-gap spectral weight, taken as the integrated intensity in the energy range around the local $E_\mathrm{F}$ for maps taken at the currents marked in Fig.~\ref{fig:1}(c).
 As expected, the in-gap intensity increases with the current throughout the sample.
 Moreover, the build-up of in-gap intensity initiates at the negative electrode (on the left hand side), consistent with the optical observation of the phase transition \cite{Zhang:2019ac,Chiriaco:2020ve}. This is clearly seen by the yellow edge developing on the left hand side of the sample in Fig.~\ref{fig:3}(a) in the panel for $I_2$. This spreads from left to right across the image as seen for $I_3$, eventually leading to the situation of $I_4$.
 Fig.~\ref{fig:3}(b) shows the spectral development at a selected point in the maps (marked with a white box in (a)), and panel (c) gives the corresponding integrated EDCs in the region of the A and B peaks.
 The evolution is consistent with that in Fig.~\ref{fig:2} and the increase of spectral weight in the gap, while small, can be observed in the EDCs.
 Fig.~\ref{fig:3}(d) and (e) give the estimated local conductivity and the applied current as a function of in-gap photoemission intensity, respectively.
 Both curves show the expected correlation between the quantities, with a clear increase of conductivity starting near the turning point of the $I/V$ curve in Fig.~\ref{fig:1}(c) and additional changes at very high currents where Joule heating is likely to play an important role in accelerating the transition.
 
 An IMT with a persistent Mott signature appears to be at odds with the conventional theoretical understanding of IMTs induced by temperature, strain, electric fields or applied currents \cite{Gorelov:2010aa,Han:2018aa,Ricco:2018aa,Bertinshaw:2019aa,Petocchi:2021wy}.
 However, a potentially important element missing in the corresponding calculations is the structural inhomogeneity accompanying the IMT.
 We show that this inhomogeneity is indeed essential for an understanding of the IMT by considering a simple model in which \ce{Ca2RuO4} structural units with a slightly different $c$-axis parameter are interfaced with each other.
 To this end, Fig.~\ref{fig:4}(a) compares the electronic density of states (DOS) for the equilibrium S-phase and a modified S$^{'}$-phase with a small (3$\%$) expansion of the $c$-axis (calculated by density function theory).
 The influence of the structural change is very small.
 Both phases are insulating with a very similar gap and the  features mainly appear shifted in energy.
 This shift, however, is crucial as it results in an effective band gap between the valence band maximum of the S-phase and the conduction band minimum of the S$^{'}$-phase that is drastically reduced compared to the gap in either of the pure phases.
 This smaller gap would be present at the interface between the two phases and results in an increased conductivity there.
 The reason for the relative energy shift is the strong covalency of the Ru 4d states combined with a distance variation between the Ru and the apical oxygen atoms that is changing the crystal field.
 As expected, the oxygen states are also affected by the modified interaction with the Ru atoms (see Fig. \ref{fig:2} and Appendix).
 
 While a single interface between the S and S$^{'}$-phases already captures the essence of the gap reduction, a more appropriate model for the experimental situation might be, for example, a superlattice with alternating S and S$^{'}$-phases, i.e., ``flat'' and ``not-so-flat'' octahedra.
 Such a structure is sketched in Fig.~\ref{fig:4}(b), constructed such that S and S$^{'}$-layers are stacked with a layer of intermediate structure in between. The calculated DOS for this superlattice is also shown in Fig.~\ref{fig:4}(b). 
Now the distinction between the contributions of the individual phases loses meaning, but the reduced gap manifests itself directly in the overall DOS. The gap reduction is qualitatively similar for other choices of the superlattice, see Appendix.
 The experimental situation is presumably more complicated than this, with inhomogeneous domain sizes that give rise to a continuum of energetically displaced and slightly different band structures.
 This will further increase the tendency for gap reduction and in-gap states, in excellent qualitative agreement with the ARPES results.
 We stress that all this proceeds without destroying the local Mott state, and that the resulting phase can be viewed as an emergent inhomogeneous band-Mott semi-metallic state.
Such a state can retain the high-energy features of the Coulomb-driven insulator while allowing for an in-gap spectral weight and a conductivity increase due to the current-driven structural inhomogeneities.
 
 In conclusion, we have shown that the current-driven IMT in \ce{Ca2RuO4} leads to the expected increase of in-gap spectral weight, but that the Mott state appears to be largely retained.
 An emergent semi-metallic state with simultaneous Mott character can be explained by the current-induced structural inhomogeneity of the sample, because the band alignment between structurally different domains effectively reduces the overall gap.
 Our work thus unveils a novel type of non-equilibrium semi-metallic state that forms at the interface of Mott domains.
 In contrast to what would be expected on the basis of well-known mechanisms, this situation arises from the structural inhomogeneity of a system, and it is entirely different than the mere coexistence of insulating and metallic domains that would be expected for a first-order phase transition.
 The creation of a novel phase from micro-structuring has not been reported in connection with the IMT in transition metal oxides, but it is somewhat reminiscent of the hidden phase in laser- or current-exposed 1T-\ce{TaS2} \cite{Stojchevska:2014aa}, where metallicity is induced in a modified Mott phase by nano-texturing \cite{Cho:2016aa}.

 \section{Methods}
 Single crystals of \ce{Ca2RuO4} were grown using the floating zone technique with Ru self-flux.
 Details about the growth procedure for the samples used in this study are reported in Ref.~\cite{Granata:2020aa}.
 The structural quality, chemical composition and electrical properties of the samples were tested by X-ray diffraction, scanning electron microscopy, energy dispersive microscopy, electron backscatter diffraction and resistivity measurements.
 The samples were found to be pure single crystals of \ce{Ca2RuO4}.
 
 Photoemission experiments were carried out at the ANTARES beamline of the SOLEIL synchrotron facility~\cite{Avila:2013aa}, using 65~eV p-polarized radiation.
 Samples were glued on an insulating (Al$_2$O$_3$) carrier using EPO-TEK H74F non-conductive epoxy and contacted from the side by DuPont 4922N silver paint with an embedded silver wire.
 They were cleaved along the $ab$ plane \textit{in situ}, in a vacuum better than 5$\times$10$^{-10}$~mbar.
 Because of this, and since the cleaving process leaves an unpredictable sample thickness, the thickness of the sample can only be estimated after the experiment.
 The dimensions of the cleaved sample in Fig.~\ref{fig:1} were measured by cutting the surrounding cured epoxy resin and exposing the sample sides.
 The dimensions were determined to be 1180~$\mu$m long by 760~$\mu$m wide by 120~$\mu$m thick.
 (The resulting cross section, which we used to estimate the current densities, is 9.12$\times$10$^{-3}$~cm$^{-2}$.)
 The image of the sample in Fig.~\ref{fig:1} was taken after the experiment.
 Lowering the current from its maximum value has a tendency to destroy the samples, and we therefore cannot exclude that certain cracks appearing in the image were not there during the measurements.
 
 The spatial resolution of the measurements was determined by scanning across a sharp feature.  For this scan, we used a lateral step size of 1~$\mu$m, and found the resolution to be 3.6~$\mu$m.
 The finite spatial resolution leads to an energy broadening of the spectra.
 This broadening is highest for the maximum local voltage drop, which is reached at a current of 2.8~mA (see Fig.~\ref{fig:1}(c)), leading to an energy broadening of 56~meV for this current.
 The overall energy resolution, excluding voltage broadening effects, was determined by measuring the Fermi edge on polycrystalline \ce{Au}:  this gave a full width at half maximum of 140~meV.
 The temperature of the sample was held constant at 200~K for the whole duration of the experiment, although we cannot exclude local Joule heating.

 The sample current was generated by a Keithley~2450.
 After every increase in current, and before measuring, the voltage was allowed to stabilize for 20 minutes.
 
 The local voltage gradient $|\nabla{V}|$ used for the calculation of the local $I/V$ curve in Fig.~\ref{fig:1}(d) was determined by tracking the fitted position of the B peak across the sample and averaging the obtained gradient values over the black square in the inset of  Fig.~\ref{fig:1}(c). 
 The local conductivity in Fig.~\ref{fig:3}(d) was estimated with the assumption of a uniform current distribution through the whole sample.
 After we found the current density $j$ and $|\nabla{V}|$, we calculated the conductivity from $\sigma = j/|\nabla{V}|$.

 We have performed density functional theory calculations using the Vienna ab-initio simulation package (VASP) \cite{Kresse93,Kresse96a,Kresse96b}.
 The core and the valence electrons were treated within the projector augmented wave \cite{Kresse99} method with a cutoff of 480~eV for the plane-wave basis.
 We have used the PBEsol exchange-correlation method \cite{Perdew08}, a revised Perdew-Burke-Ernzerhof (PBE) that improves equilibrium properties of solids.
 PBEsol+$U$ is the approach that we have followed to take into account the correlations associated with the Ru 4d states.
 We have considered $U$=3 eV for the antiferromagnetic insulating phase of ruthenates \cite{Autieri_2016,Gorelov:2010aa}, and regarding the Hund coupling we assume
 $J_H$ = 0.15 $U$ in agreement with the approaches based on the constrained
 random phase approximation for 4d-electrons \cite{Vaugier12}.
 The values of the lattice constants are $a_S$=5.3945~{\AA}, $b_S$=5.5999~{\AA}, $c_S$=11.7653~{\AA} in the S-Pbca phase.
 For the S$^{'}$-Pbca phase we assume a variation of three percent for the $c_{S^{'}}$ axis length with respect to the S-phase, \emph{i.e.}, $\frac{(c_{S^{'}}-c_S)}{c_S}=0.03$.

\section{Appendix}

\subsection{X-ray diffraction data}

To supplement the ARPES data, X-ray diffraction (XRD) measurements were performed for samples from the same batch in different conductivity regimes. As in the ARPES experiments, the temperature was set to $T$ = 198~K and a d.c. current was applied in the $a$-$b$ plane of the sample. XRD patterns were obtained in reflection by an automatic Bruker D8 Advance diffractometer equipped with a non-ambient stage and by using the nickel filtered Cu-K$_\alpha$ radiation. This leaves the contributions of the K$_{\alpha1}$ and K$_{\alpha2}$ doublet in the data. 

\begin{figure}
  \includegraphics[width=0.4\textwidth]{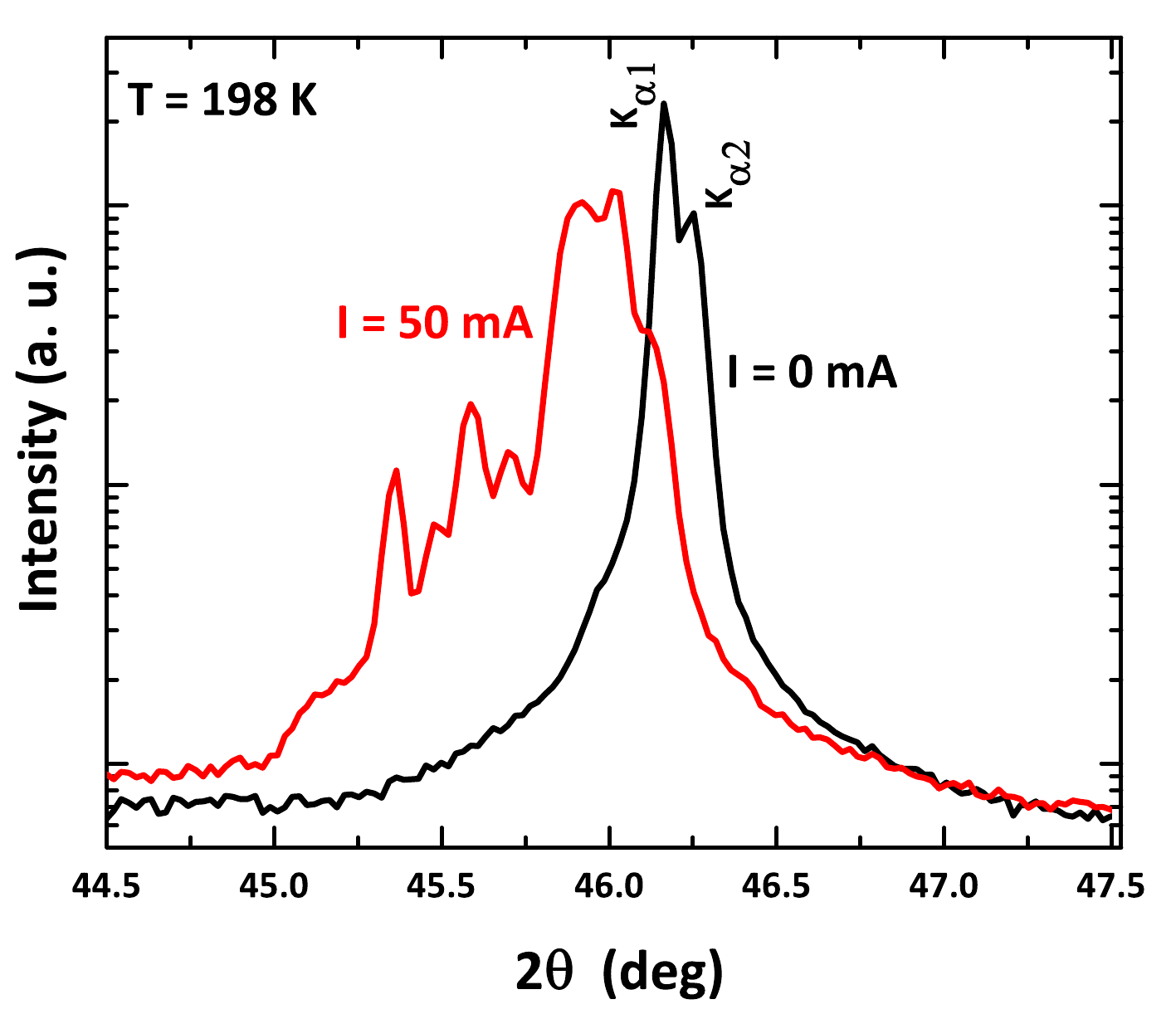}\\
  
  \caption{X-ray diffraction scans around the (006) reflection performed at $T$ = 198~K for $I$ = 0~mA and $I$ = 50~mA. The simultaneous presence of  Cu K$_{\alpha1}$ and K$_{\alpha2}$ radiation leads to a characteristic double peak structure (indicated).
  }
  \label{fig:S11}
\end{figure}

In Fig.~\ref{fig:S11}, two representative XRD scans are shown, to allow the comparison between static ($I$ =~0 mA) and the \emph{in-operando} conditions ($I$ =~50 mA)---i.e., for a high current in the region where a negative differential resistance is measured in the $I$-$V$ curve. We calculated $c$-axes values according to the Bragg law, by following the position of the (006) Bragg reflection. The $c$-axis value of the insulating S phase at $I$ = 0~mA is $c$ = 11.789(4)~\AA. This value is in agreement with the one reported in literature at the same temperature \cite{Friedt:2001aa}. When current is applied, lattice distortions occur and multiple $c$-axes related to the emergent different domains are detected. This is clearly evident in Fig.~\ref{fig:S11} because the diffraction peak does not only shift in angle upon the application of the current. It also broadens considerably, indicating a quasi continuum of domains with different $c$-axis lengths in the region probed. This formation of domains with different $c$-axis has also been observed by applying a current along $c$-axis values \cite{Cirillo:2019aa}. In general, XRD measurements allow to infer that different crystallographic domains with $c$-parameters in the insulating region coexist in a wide range of applied current values.

\subsection{Data tracking the IMT}

\begin{figure}
  \includegraphics[width=0.5\textwidth]{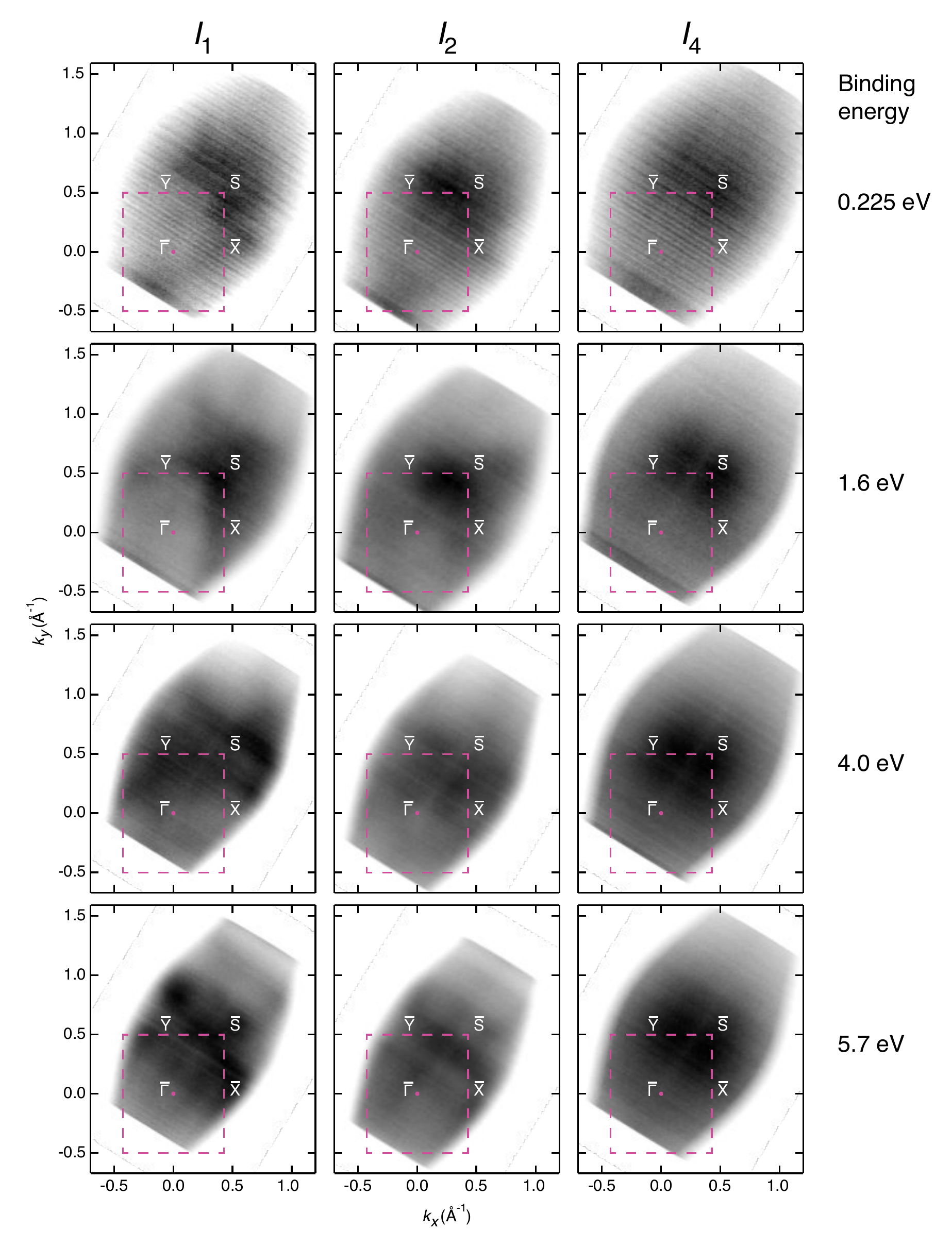}\\
  \caption{Constant energy cuts supplementing the data in Fig.~\ref{fig:2}(a) - (c) of the main text, at selected binding energies and at currents $I_1$, $I_2$, and $I_4$.
  }
  \label{fig:S9}
\end{figure}

\begin{figure}
  \includegraphics[width=0.5\textwidth]{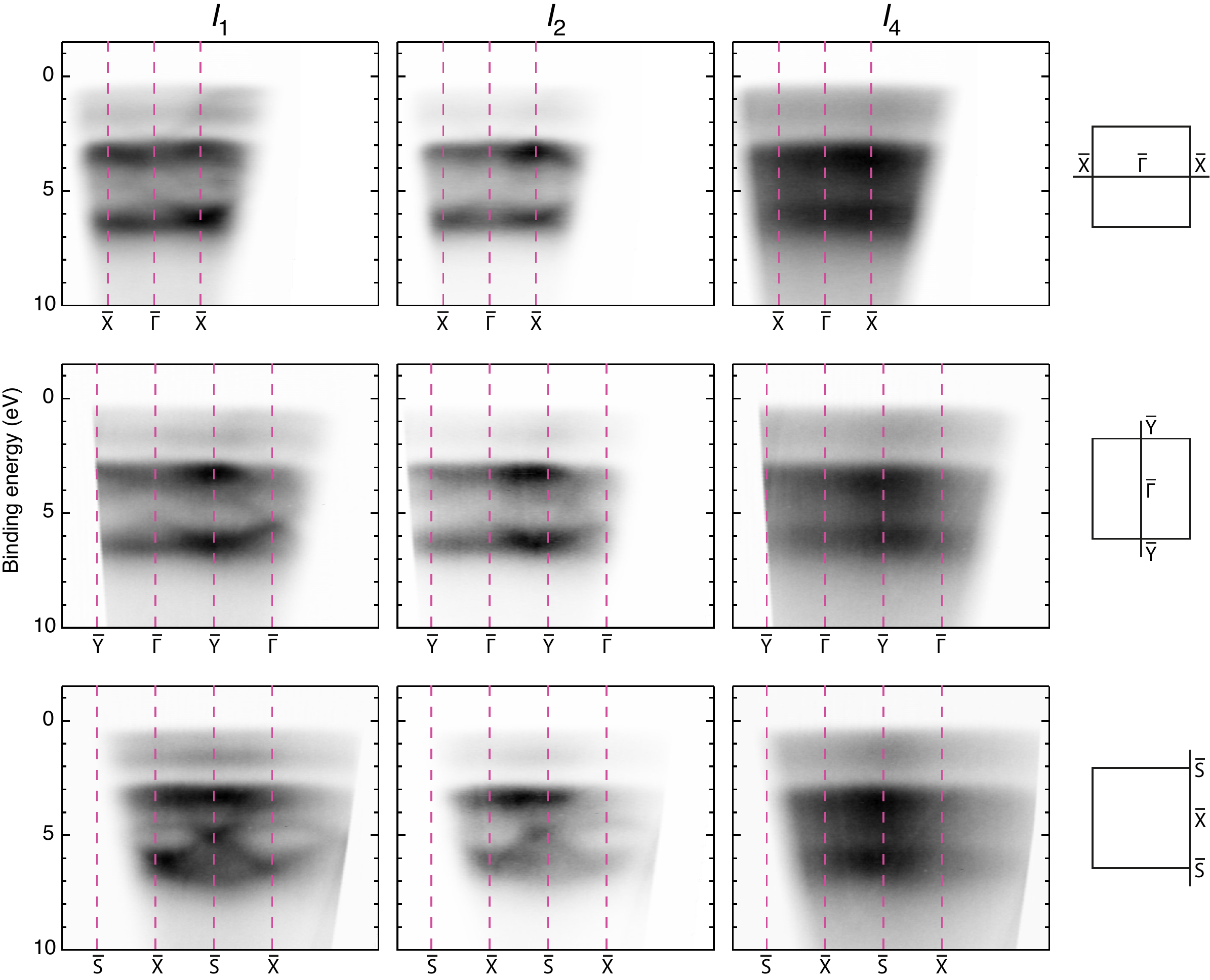}\\
  \caption{Photoemission along selected high symmetry directions supplementing the data in Fig.~\ref{fig:2}(d)-(f) of the main text at currents $I_1$, $I_2$, and $I_4$.
  }
  \label{fig:S10}
\end{figure}

In this section, we provide additional data on the IMT as characterised in Fig.~\ref{fig:2} and \ref{fig:3} of the main text.
Figures \ref{fig:S9} and \ref{fig:S10} show the same kind of photoemission intensity maps at constant binding energy and as a function of binding energy and $k$ as given in Fig.~\ref{fig:2}, but for more binding energies and directions in reciprocal space.
This additional data is consistent with the trend described in connection with Fig.~\ref{fig:2}.
For higher currents, we find an increasing $k$-broadening of the photoemission features.

\begin{figure*}
  \includegraphics[width=0.7\textwidth]{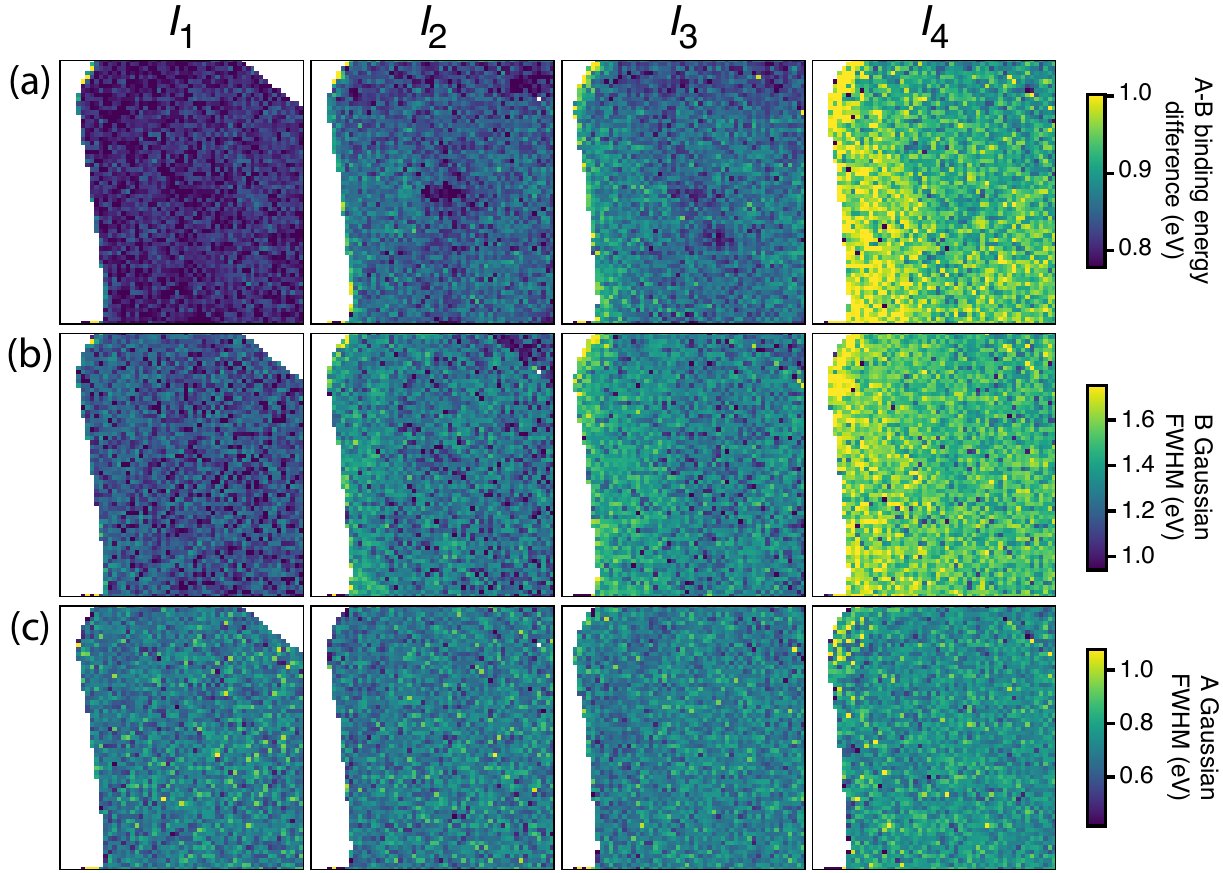}\\
  \caption{Maps showing the evolution of fit parameters for the A and B peaks, complementing the results in Fig.~\ref{fig:3} of the main text.
    (a) Binding energy difference between the A and the B peaks.
    (b) Gaussian FWHM for the B peak.
    (c) Gaussian FWHM for the A peak.
  }
  \label{fig:S3}
\end{figure*}

Additional maps characterising the IMT mapped across the sample are given in Fig.~\ref{fig:S3} as a supplement to Fig.~\ref{fig:3} in the main text and to illustrate the behaviour of the parameters needed to fit the photoemission intensity at low binding energy to the A and B peaks.
Fig.~\ref{fig:S3}(a) displays the spatial dependence of the binding energy splitting between the A and B peaks.
This shows a marked increase with current, by roughly 130~meV.
The splitting has some degree of spatial dependence, with the left hand side closest to the negative electrode showing a slightly larger splitting.
Figures~\ref{fig:S3}(b) and (c) show the spatial dependence of the Gaussian full width at half maximum (FWHM) for the B and A peak, respectively.
Both increase at higher currents but the broadening is stronger for the B peak.
The A peak shows a rather uniform broadening across the sample while the B peak shows a small indication that the broadening is larger closer to the negative contact.
The increased in-gap photoemission intensity thus results from a combination of the increased FWHM of the A peak and the larger separation between the B and the A peak.

\subsection{Additional data from an experiment at MAX IV and comparison to SOLEIL results}

\begin{figure*}
  \includegraphics[width=0.7\textwidth]{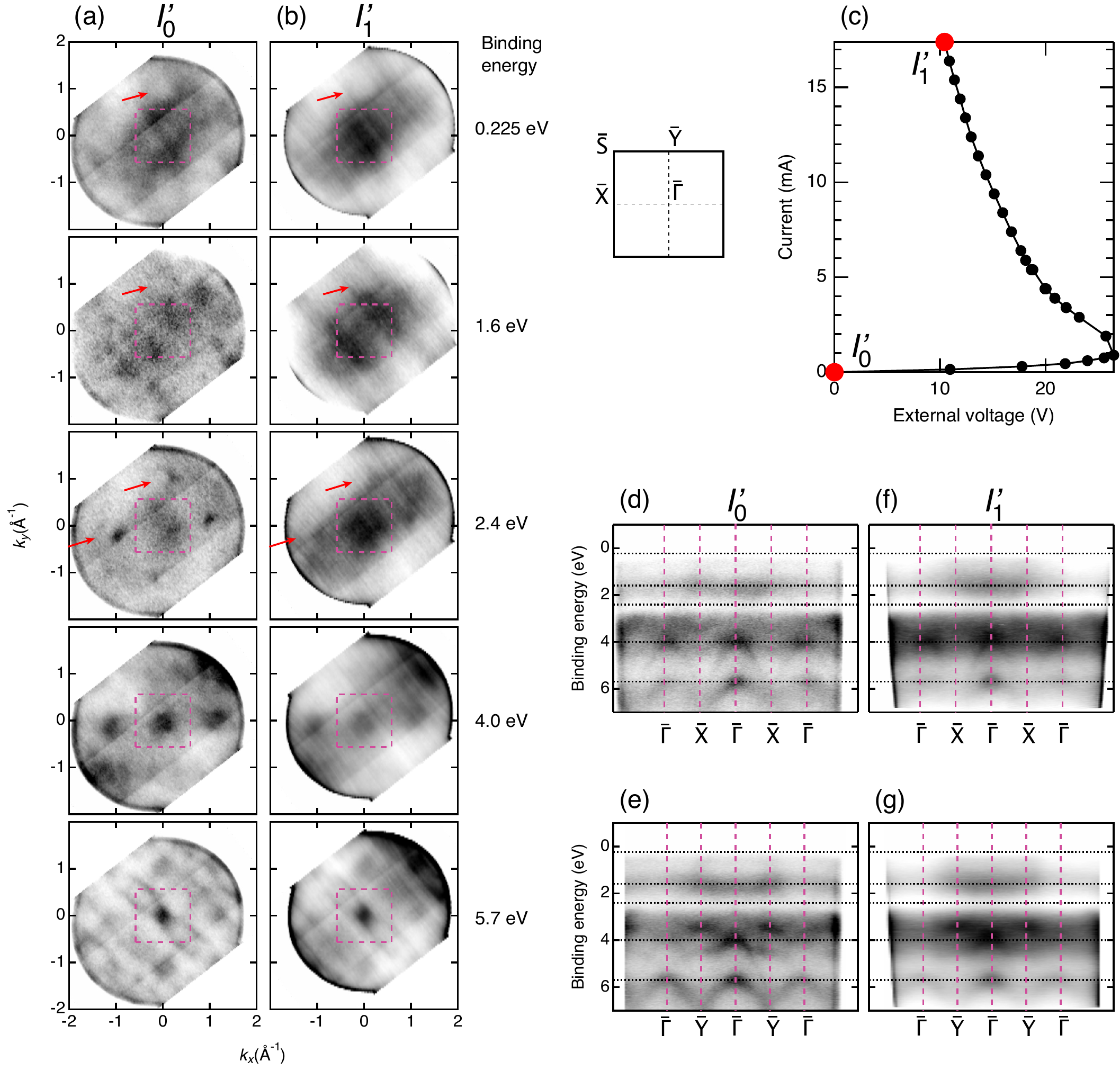}\\
  \caption{ARPES angle scans acquired at MAX IV.
    (a), (b) Photoemission intensity at constant binding energy for $I_{0}^{\prime}=0$ and for the highest current $I_{1}^{\prime}$.
    Red arrows mark the circular features associated with the dispersing \ce{Ru} $d_{xy}$ states.
    Yellow dashed rectangles mark a single Brillouin zone.
    (c) Externally measured $I/V$ curve.
    (d)-(g) Photoemission intensity along high symmetry directions for the data set with no current (panels (d) and (e)) and $I_{1}^{\prime}$ (panels (f) and (g)).
    The dashed horizontal lines correspond to the energies of the data in panels (a) and (b).
  }
  \label{fig:S1}
\end{figure*}

A similar experiment to what is reported in the main text was performed at the Bloch beamline of the MAX IV laboratory, under the same conditions and with consistent results.
In this section, we present the data from MAX IV and make a comparison to the data taken at SOLEIL.
At MAX IV, ARPES measurements were performed using a photon energy of 148~eV and s-type polarization.
Due to this higher photon energy, the MAX IV data covers a larger $k_{\parallel}$ range than the SOLEIL results.
The different photon energy also leads to changes in the photoemission matrix elements such that the same structures have different relative intensities.
The light spot size at MAX IV was larger than at SOLEIL (between 10~$\mu$m and 20~$\mu$m), resulting in a still moderate field-induced energy broadening.
An estimate along the lines discussed in the main text finds this broadening to be smaller than 15~meV for $I_{1}^{\prime}$ (see Fig.~\ref{fig:S1}).

Equilibrium data from MAX IV at $I_{0}^{\prime}=0$, and data acquired for the highest applied current of $I_{1}^{\prime}=$17.4~mA, is reported in Fig.~\ref{fig:S1}, along with the externally measured two-point $I/V$ curve.
The current-induced changes are very similar to those reported in the main paper.
The current primarily leads to a $k$-broadening of the features in the oxygen bands.
A small change in the \ce{Ru} $d_{zx}$ and $d_{yz}$ bands is also observed.
Most importantly, the essential characteristics of the Mott state, as manifested in the non-dispersive A and B bands, is retained, even at the very high current $I_{1}^{\prime}$.

Note that the overall broadening at the highest current $I_{1}^{\prime}$ appears to be less pronounced than at the highest current in the SOLEIL data set $I_4$ where the bands are completely washed out.
The reason for this is that $I_4$ is a substantially higher current than $I_{1}^{\prime}$ with respect to the turning point of the $I/V$ curve.
At MAX IV it was not possible to reach such a high current before the sample was destroyed by the structural changes accompanying the phase transition.

\begin{figure}
  \includegraphics[width=0.3\textwidth]{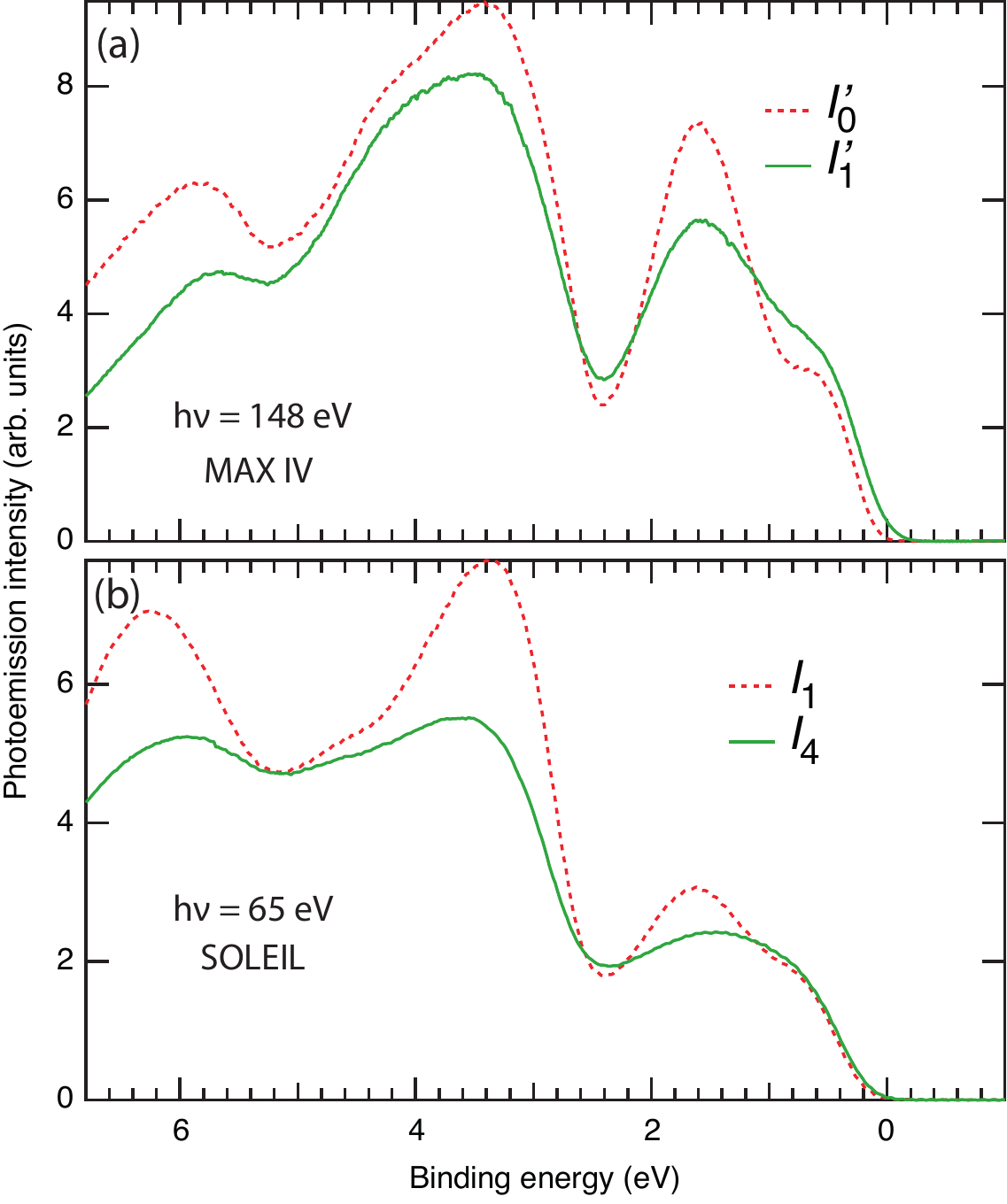}\\
  \caption{Energy distribution curves from two different experiments.
    (a) Data from MAX IV, extracted from Fig.~\ref{fig:S1}.
    (b) Data acquired at SOLEIL from the measurements described in the main text.
    In both cases, the momentum integration range corresponds to a full Brillouin zone.
  }
  \label{fig:S2}
\end{figure}

The increase of in-gap intensity signalling the IMT for high currents is also seen in the data from MAX IV.
As in the main text, this is best shown by integrating the angle-resolved data to give one-dimensional EDCs and aligning the energy scale to the position of the B peak obtained from a fit.
Such EDCs are shown in Fig.~\ref{fig:S2} and compared to the corresponding data from SOLEIL.
The increase of photoemission intensity at $E_\mathrm{F}$ is even more evident in the MAX IV data, but we stress again that the line shapes and relative intensities of the peaks are not directly comparable, due to the different photon energies in the two experiments.

\begin{figure}
  \includegraphics[width=0.3\textwidth]{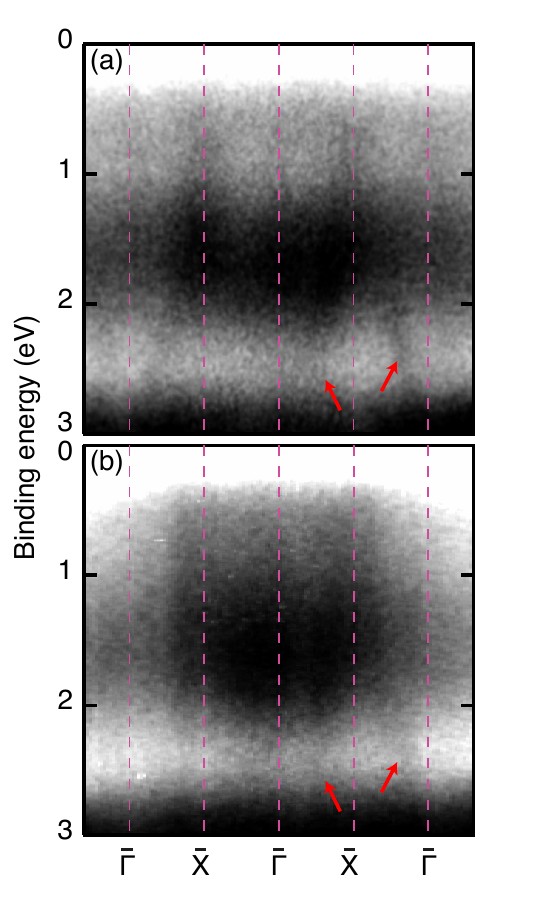}\\
  \caption{
    Data from MAX IV showing a zoomed in portion of the data displayed in Fig.~\ref{fig:S1}.
    In (a), data is taken at no current, while in (b) data is taken at $I_{1}^{\prime}$.
    The red arrows mark some of the dispersive features discussed in the text.
  }
  \label{fig:S6}
\end{figure}

\begin{figure}
  \includegraphics[width=0.3\textwidth]{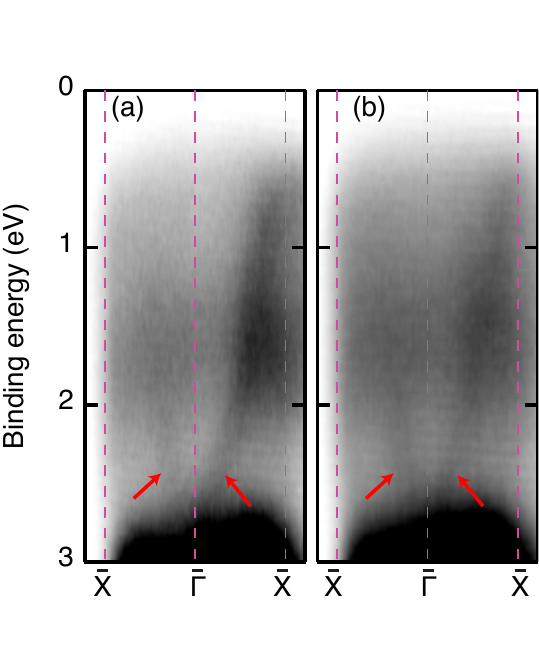}\\
  \caption{
    Data from SOLEIL showing a zoomed in portion of the data displayed in Fig.~\ref{fig:2}.
    In (a), data is taken at $I_1$, while in (b) data is taken at $I_2$.
    The red arrows mark the dispersive features discussed in the text.
  }
  \label{fig:S7}
\end{figure}

We conclude with a comparison of the dispersive \ce{Ru} $d_{xy}$ band between the data sets from MAX IV and SOLEIL.
This band cannot only be seen in the constant energy surfaces (circular features marked by red arrows in Fig.~\ref{fig:2} of the main text and Fig.~\ref{fig:S1}) but also as dispersive features along high symmetry directions of the Brillouin zone, similar as in published data \cite{Sutter:2017aa}.
We show this by zoomed-in versions from the data in Fig.~\ref{fig:S1} (MAX IV) and Fig.~\ref{fig:2} of the main text (SOLEIL) in Figures~\ref{fig:S6} and \ref{fig:S7}, respectively.
The figures show both a low and a high current regime.
The $d_{xy}$ bands experience some broadening but are not washed out.

\clearpage
\subsection{Data of electronic structure and superlattice density of states}

\begin{figure}
 \includegraphics[angle=270,width=0.505\textwidth]{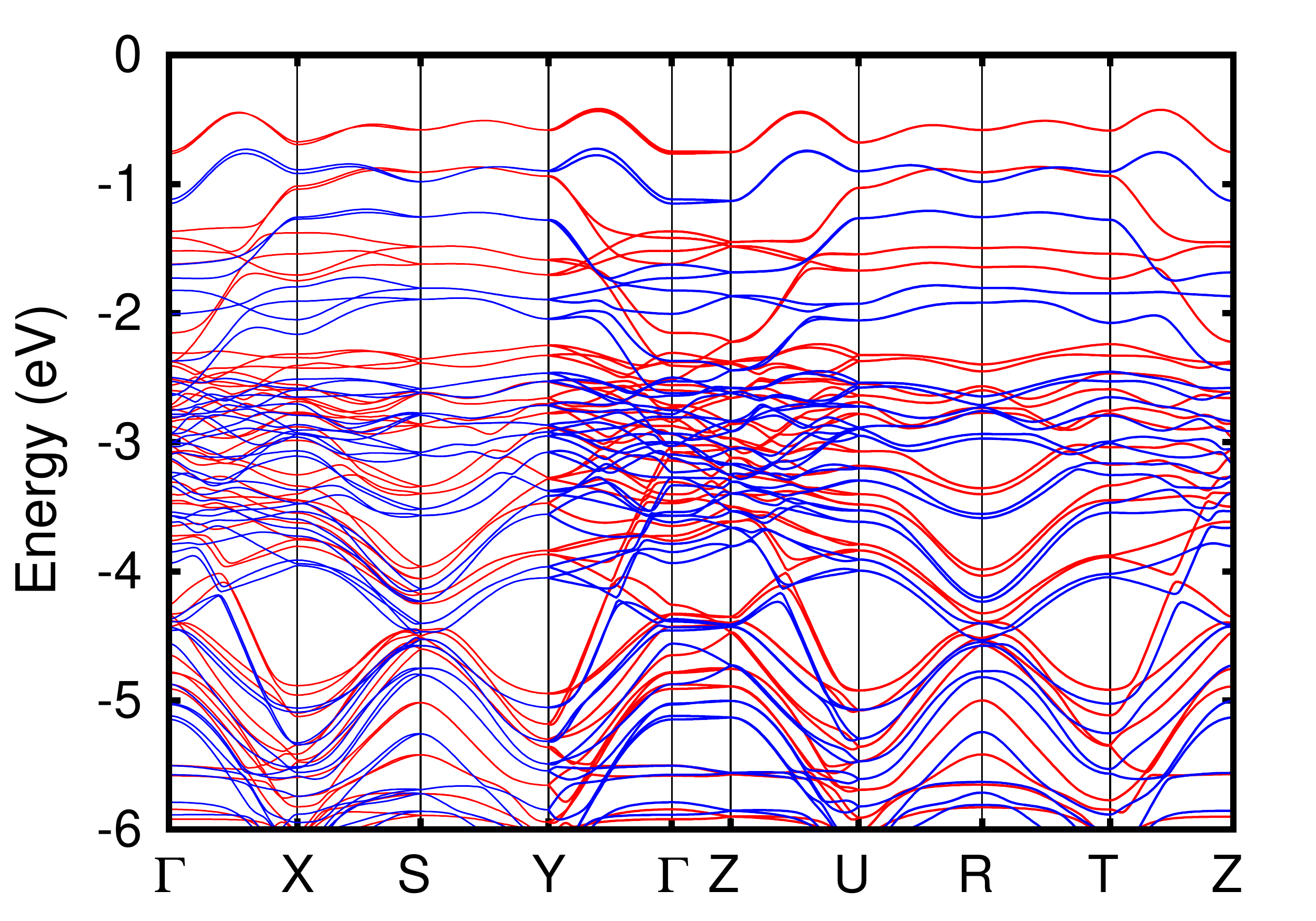}\\
  \caption{Band structure of the bulk Ca$_2$RuO$_4$ for the S (red line) and the S$^{'}$ (blue line) phases. The zero of the energy scale is fixed to the valence band maximum of the superlattice in Fig.~\ref{fig:4}(b) of the main text.
  }
  \label{fig:T3}
\end{figure}

The electronic band structures for the S- and S$^{'}$-phases are reported in Fig.~\ref{fig:T3}. In this representation of the electronic structure, it becomes evident how even a small energy shift in the region of the oxygen bands will manifest itself as a $k$-broadening in the ARPES data.

\begin{figure}
\includegraphics[angle=0,width=0.495\textwidth]{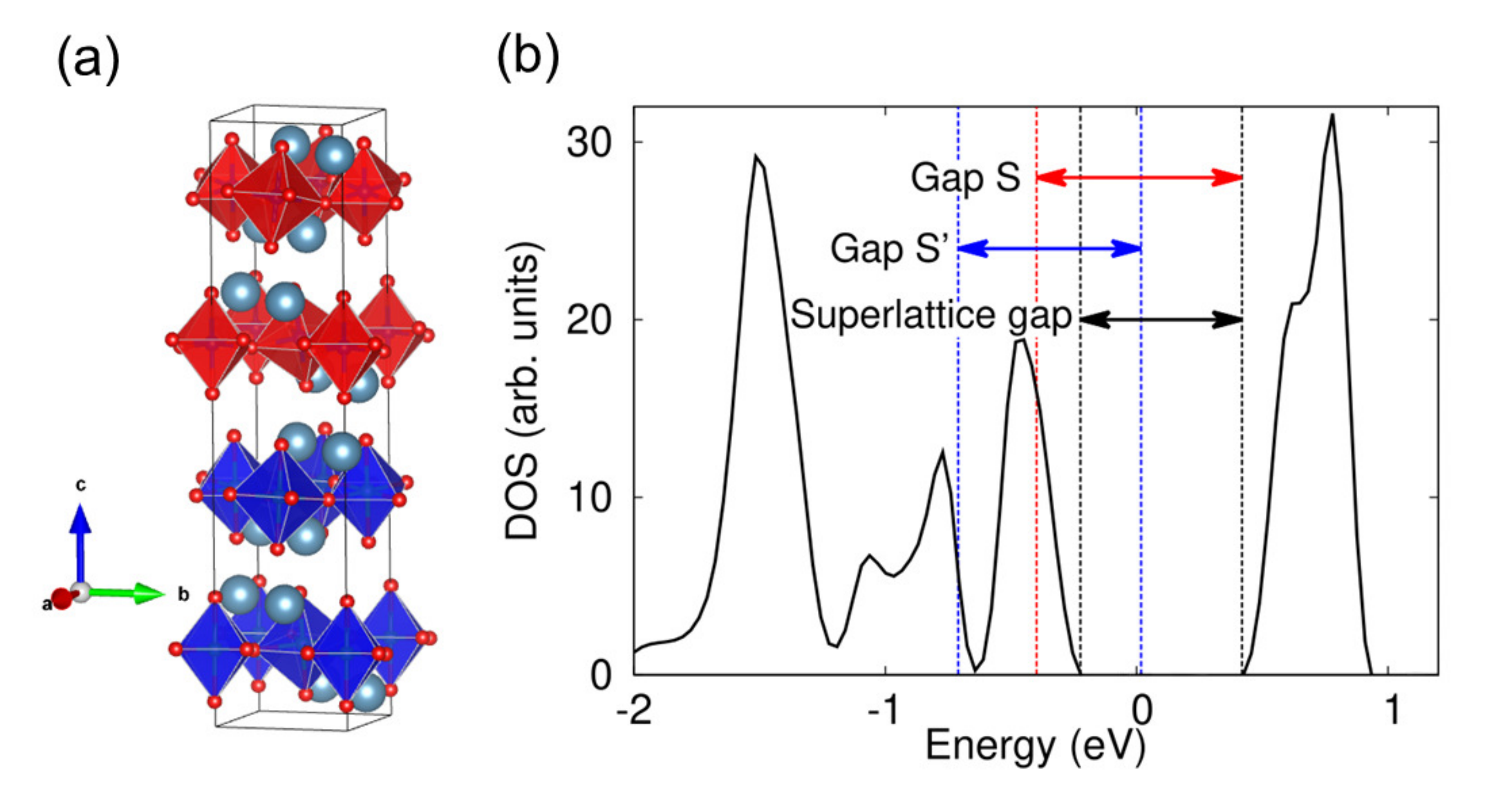}\\
  \caption{(a) Crystal structure of a superlattice formed by two layers of S and two layers of S$^{'}$ unit cells with a sharp interface. (b) Total density of states for this superlattice. The zero of the energy scale is fixed to the valence band maximum of the other superlattice in Fig.~\ref{fig:4}(b) of the main text. 
  }
  \label{fig:T5}
\end{figure}

We also evaluate other combinations of stacking of the S and S$^{'}$-phases as compared to the case presented in the main text. For each examined configuration, we find a reduction of the band gap. In Fig. \ref{fig:T5}(a), we show a representative crystal structure of a supercell with a stacking of two S and two S$^{'}$ layers and a sharp interface between S and S'-phase. The corresponding DOS is presented in Fig. \ref{fig:T5}(b) showing a reduction of the band gap in the superlattice as compared to the bulk phases (see Fig.~4 (a) in the main text). The gaps of the S and S'-phases are 0.84 and 0.75~eV, respectively.
The gap of the superlattice with a sharp interface between S and S' phase is 0.68 eV, while in case of the non-sharp interface the gap can be lower. Indeed, the superlattice with rough interface presented in Fig.~4 of the main text has a gap of 0.35~eV. 

\vspace{1mm}

\begin{acknowledgments}
This work was supported by VILLUM FONDEN via the Centre of Excellence for Dirac Materials (Grant No. 11744). G. C. and C. A. are supported by the Foundation for Polish Science through the International Research Agendas program co-financed by the
European Union within the Smart Growth Operational Programme (Grant No. MAB/2017/1).
G. C. and C. A. acknowledge the access to the computing facilities of the Interdisciplinary
Center of Modeling at the University of Warsaw, Grant G84-0, GB84-1 and GB84-7. We gratefully acknowledge Dr. Guerino Avallone for performing the X-ray scattering measurements. We acknowledge SOLEIL for provision of synchrotron radiation facilities and MAX IV Laboratory for time on Beamline Bloch under Proposal 20190681. Research conducted at MAX IV, a Swedish national user facility, is supported by the Swedish Research council under contract 2018-07152, the Swedish Governmental Agency for Innovation Systems under contract 2018-04969, and Formas under contract 2019-02496.
\end{acknowledgments}

\end{document}